# Time-Ordered Product Expansions

# for Computational Stochastic Systems Biology

Eric Mjolsness

UC Irvine Department of Computer Science

emj@uci.edu

September 2012

**Abstract**

The time-ordered product framework of quantum field theory can also be used to understand salient phenomena in stochastic biochemical networks. It is used here to derive Gillespie's Stochastic Simulation Algorithm (SSA) for chemical reaction networks; consequently, the SSA can be interpreted in terms of Feynman diagrams. It is also used here to derive other, more general simulation and parameter-learning algorithms including simulation algorithms for networks of stochastic reaction-like processes operating on parameterized objects, and also hybrid stochastic reaction/differential equation models in which systems of ordinary differential equations evolve the parameters of objects that can also undergo stochastic reactions. Thus, the time-ordered product expansion (TOPE) can be used systematically to derive simulation and parameter-fitting algorithms for stochastic systems.

## 1    Introduction

The master equation for a continuous-time stochastic dynamical system may be expressed as $d\,p/d\,t = W \cdot p$ where $W$ is the time-evolution operator, often an infinite-dimensional matrix. Particular choices for $W$ lead to the special case of the "chemical master equation" for stochastic chemical kinetics, often useful in bioligical applications; we will see this and other applications below. The general master equation has the formal solution $p(t) = \exp(t\,W) \cdot p(0)$. If $W$ can be decomposed as a sum $W_0 + W_1$, then there is a perturbation theory for $\exp(t\,W)$ in terms of $\exp(t\,W_0)$ and its perturbations by $W_1$. The time-ordered product expansion (which we refer to by the acronym TOPE) gives a formula for the solution of a master equation [1-3] which can be expressed as follows [4]:

$$\exp(t\,W) \cdot p_0 = \exp(t\,(W_0 + W_1)) \cdot p_0$$

$$= \sum_{k=0}^{\infty} \left[ \int_0^t d\,t_k \int_0^{t_k} d\,t_{k-1} \cdots \int_0^{t_2} d\,t_1 \exp((t-t_k)\,W_0) \cdot W_1 \cdot \exp((t_k - t_{k-1})\,W_0) \cdots \cdot W_1 \cdot \exp(t_1\,W_0) \cdot p_0 \right] \quad (1)$$

This expression can be derived (as in [4]) by expanding in powers of $W_1$, each expanded to all orders in $W_0$, and using the normalization formula for the Dirichlet distribution to subdivide the time interval $[0, t]$ into $k$ subintervals.

Since $W_0$ and $W_1$ do not generally commute, this expression involves alternation from right to left of $W_0$ and $W_1$ related operations. Using the "time-ordered exponential" of operators [5], this result can be compactly reexpressed as:

$$\exp(t\,(W_0 + W_1)) = \exp(t\,W_0)\left(\exp\left(\int_0^t W_1\,(\tau)\,d\,\tau\right)\right)_+ \quad (2)$$

where

$$W_1\,(\tau\,) \equiv \exp(-\,\tau\,W_0)\,W_1\,\exp(\tau\,W_0)\,. \tag{3}$$

Here $\left(\exp\!\left(\int_0^t G(\tau)\,d\tau\right)\right)_+$ is obtained term by term from the Taylor series for the operator exponential, by reordering all monomials containing terms evaluated at different times so that they are indexed by ordered sequences of times $(\tau_k,\,...,\,\tau_1)$ that increase right to left (details reviewed in Section 3.5.1 below). In field theory it is standard to prefer Equation 2 over Equation 1 for theoretical calculations, but for algorithmic concreteness this paper will favor the more explicit expression Equation 1 where possible. Indeed, each summand of Equation 1 already looks like a matrix or operator product operation "·". which sum over states, is supplemented by integration over an extra time variable. This observation will be made precise in Section 3. In general there is a risk that the infinite sum over terms could diverge. However, the master equation must conserve total probability and this constrains $W$ to have zero column-sums and also constrains the spectrum of $W$ to have nonpositive real parts. In this setting some decompositions $W = W_0 + W_1$ converge well enough, as we will see by example below.

One particular specialization of TOPE lets us derive Gillespie's Stochastic Simulation Algorithm (SSA): take $W_0 = -D =$ the diagonal part of $W$, and $W_1 = \hat{W} =$ the off-diagonal part of $W$. Then for chemical reaction networks TOPE generates Feynman-like diagrams. An example is illustrated below for the simple reaction network with just two reactions, the forwards and backwards parts of the generic trivalent reaction $A + B \rightleftharpoons C$, to which others can be reduced.

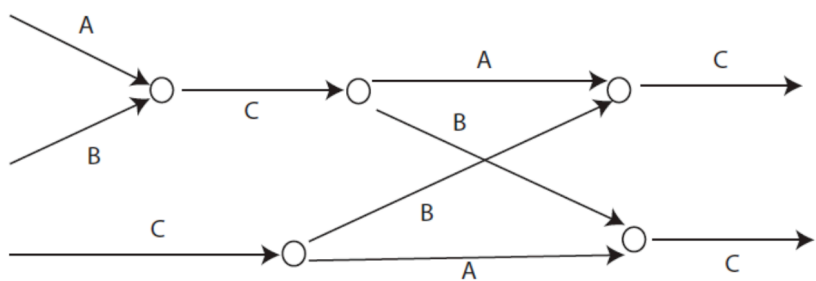

Figure 1. A time history of the reaction $A + B \rightleftharpoons C$. Time flows left to right. Open circles represent reaction events, with probability factor $\times W_1$. In between reaction events are unimolecular particle propagators $\exp((t_k - t_{k-1})\,W_0)$, labelled by arrows and particle names (repeated for clarity). This is a non-spatial version of the Lee model in quantum field theory (cf. for example [6]).

The TOPE (Equation 1 or Equation 2) can be applied recursively, since it reduces one operator exponential $\exp(t\,W)$ to another one $\exp(t\,W_0)$. This fact will be exploited in Section 3 below. But eventually one must get to an operator exponential that is tractable by other means. One way to do this is to let $W_0 = D =$ the diagonal part of $W$, as in the SSA algorithm derivation below.

# 2    Methods

## 2.1    Creation/annihilation operator notation

We will use operator notation for molecule (or other reactant) creation and annihilation state changes [1-4]. Here we just review the notation as used in [4]. The elementary operators $a$ and $\hat{a}$ act (respectively) to destroy and create identical particles of a given type. In the particle-number basis their elements have the entirely off-diagonal expressions

$$a_{ij} = j\,\delta_{i\,j-1} \quad \text{and} \quad \hat{a}_{ij} = \delta_{i\,j+1} \quad \text{for all } i, j \text{ in } \{0, 1, 2, \ldots\}\,. \tag{4}$$

Here $\delta_{ij}$ is the Kronecker delta function. The creation and annihilation operators satisfy the Heisenberg algebra $[a, \hat{a}] = I$ but are different from those of quantum mechanics because they are not conjugates or transposes of one another. (This is the reason we do not denote the creation operator $a^{\dagger}$, as it is in quantum mechanics, or $a^{*}$ .) Instead of being conjugate to $\hat{a}$, the annihilator $a$ encodes the chemical law of mass action since its nonzero entries are equal to the number of particles available to react or decay. The diagonal "number operator" is $N \equiv \hat{a}\,a$ .

The creation and annihilation operators may be represented in terms of their action on probability generating functions $g(z) = \sum_{n=0}^{\infty} p_n\,z^n$, where $p_n$ is the probability that there exist $n$ particles of a given type. In this case:

$$a = \partial_z \cdots \quad \text{and} \quad \hat{a} = z \times \cdots \tag{5}$$

In the presence of different types of particles (eg. molecules or other objects) the creation/annihilation operator notation is generalized, eg to $a_\alpha$ and $\hat{a}_\beta$ for molecule types $A_\alpha$ , in which all operators for unequal types commute:

$$[a_\alpha, \hat{a}_\beta] \equiv a_\alpha\,\hat{a}_\beta - \hat{a}_\beta\,a_\alpha = I\,\delta_{\alpha\beta}$$

Operating on an empty "vacuum" state $|0\rangle$ with no objects, the monomials in the creation operators $\hat{a}_\beta$ span a Fock space. Molecule or object types indexed by $\alpha$ may even be taken to include arbitrary discrete-valued molecular attributes (or attributes of other objects) such as phosphorylation state or integer-valued parameters. Continuous-valued parameters such as position (in quantum field theory it would more naturally be the conserved momentum, unlike the typical viscous-medium dynamics in biology) may be encoded into a real-valued vector argument $x$ which requires a Dirac delta function instead of a Kronecker delta function, so for example:

$$[a_\alpha(x), \hat{a}_\beta(y)] = I\,\delta_{\alpha\beta}\,\delta(x-y) \tag{6}$$

A non-molecular example of such parameterized objects would be: cells of a given real-valued volume or lengthscale as in  Section 3.5.5 below.

However for some attributes such as real-valued object positions one may wish to limit the state space to between zero or $n_{max,\alpha}$ molecules (or other objects) at each unique real value. The resulting commutator is still diagonal as described in [4]. The particular case $n_{max,\alpha} = 1$ is *not* a stochastic version of fermions because particles with different types or values of the attributes still commute rather than anticommuting.

The basic rule for translating chemical reactions into creation/annihilation operator notation is: first, annihilate all objects on the incoming or left hand side of a reaction; then create all the objects on the outgoing or right hand side of a reaction. Thus, the off-diagonal part of the operator for a reaction

$$\{A_{\alpha(p)}(p) \mid 1 \le p \le p_{max}\}_* \longrightarrow \{A_{\beta(q)}(q) \mid 1 \le q \le q_{max}\}_* \quad \textbf{with} \quad \text{reaction rate} \ \ \rho_r$$

that converts an incoming multiset $\{\cdots\}_*$ of numerically parameterized reactants $\{A_{\alpha(p)}(p) \mid p \in \text{lhs}(r)\}_*$ (reactants can appear multiple times in a multiset) into an outgoing multiset $\{A_{\beta(q)}(q) \mid q \in \text{rhs}(r)\}_*$, is:

$$\hat{O}_r = \rho_r \left[ \prod_{q \in \text{rhs}(r)} \hat{a}_{\beta(q)}(y_q) \right] \left[ \prod_{p \in \text{lhs}(r)} a_{\alpha(p)}(x_q) \right]. \tag{7}$$

There is one such operator for every possible set of values for the numerical parameters. Since time-evolution operators for different processes just add, a generic operator for all parameter values must sum and/or integrate the operator of Equation 7 over all the parameters, in the Cartesian product of measure spaces in which they take values. Equation 7 adds probability to the new state of the system, but does not take it away from the old state of the system before a reaction. That job requires a negative diagonal matrix as shown in Equation 9 below. In the case of Equation 7, the corresponding diagonal operator is $D_r = \rho_r \left[ \prod_{a \in \text{lhs}(r)} N_{\alpha(p)}(x_p) \right]$. Examples are provided in [4] and below.

## 2.2    Solvable example: An exact solution for SSA behavior

For a few very simple examples, we can not only solve analytically for the behavior of the biochemical system, but we can even add in the behavior of the SSA simulation algorithm and solve for that exactly as well. For example consider the minimal bidirectional reaction $A \longleftrightarrow \varnothing$. This case is analytically solvable, including the complete statistics of its SSA algorithm simulation. It has forward synthesis and backwards decay reactions. The operator expression is therefore:

$$W = k_s(\alpha \hat{a} - I) + k_d(\alpha a - N) \tag{8}$$

Here $\alpha = 1$ is the generating function variable for the total number of reactions, corresponding to off-diagonal matrix elements of $W$. Power series in $\alpha$ will decompose total probability according to this number.

Translating the master equation for Equation 8 into a PDE in the two variables $t$ and $z$ using representation Equation 5, and solving analytically, this model has the exact solution

$$g_m(z, t \mid \alpha) = \left(\alpha + (z - \alpha) e^{-k_d t}\right)^m \exp\left[-\frac{k_s}{k_d}\left((1 - \alpha) k_d t + \left(z\alpha - \alpha^2\right)\left(e^{-k_d t} - 1\right)\right)\right]$$

$$= \left(\alpha + (z - \alpha) e^{-k_d t}\right)^m \exp\left[\frac{k_s}{k_d}\left(z\alpha - 1\right)\left(1 - e^{-k_d t}\right)\right]$$

$$\times \exp\left[\frac{k_s}{k_d}\left(\alpha^2 - 1\right)\left(k_d t + e^{-k_d t} - 1\right)\right]$$

= Binomial initial condition with decay $*$ Poisson on forward reactions

$*$ Poisson on forward / backward reaction pairs

As usual $z$ is the generating function variable whose exponent is the total number $n_A$ of $A$ molecules or particles, $m$ is the initial number of molecules, and $t$ is continuous time. The $*$ operation is a convolution of probability distributions. A product of generating functions with the same variable is a convolution of distributions [7]. Note the interpretation in terms of Binomials and Poissons with time-varying parameters. The third factor represents a linearly increasing number of canceling forward/backward reaction pairs as a function of time - a kind of random walk.

The full derivation below will generalize this solvable example, again separating the diagonal from the off-diagonal terms in $W$.

## 2.3 Notation for SSA rederivation from TOPE

One specialization of TOPE lets us derive SSA for biochemical reaction networks, as follows. First decompose $W$ into nonegative off-diagonal and nonpositive diagonal parts, as must be possible by the conservation and nonnegativity of probability. For example conservation of probability implies $\forall p \; 0 = d(\mathbf{1} \cdot p)/dt = (\mathbf{1} \cdot W) \cdot p \Rightarrow \mathbf{1} \cdot W = 0$. Then

$$W = \hat{W} - D, \quad \text{where} \quad D = \mathrm{diag}(\mathbf{1} \cdot \hat{W}), \quad \text{i.e.}$$
$$\hat{W}_{IJ} = (1 - \delta_{IJ}) W_{IJ} \quad \text{and} \quad D_{IJ} = \delta_{IJ} \sum_K \hat{W}_{KJ} \tag{9}$$

where $I$ and $J$ index the possible states of the system. To prevent negative probabilities from evolving under the master equation, all entries of $\hat{W}$ and therefore $D$ must be nonnegative. In this circumstance the TOPE becomes:

$$\exp\left(t\left(\hat{W} - D\right)\right) =$$

$$\sum_{k=0}^{\infty} \left[ \int_0^t \cdots \int_0^t \left(\Pi_{q=0}^k \, d\tau_q\right) \delta\left(t - \sum_{q=0}^k \tau_q\right) \exp(-\tau_k D) \, \hat{W} \cdots \exp(-\tau_1 D) \, \hat{W} \exp(-\tau_0 D) \right]$$

$$= \sum_{k=0}^{\infty} \int_0^t \cdots \int_0^t \left(\Pi_{q=0}^k \, d\tau_q\right) \delta\left(t - \sum_{q=0}^k \tau_q\right) \exp(-\tau_k D) \left[ \prod_{q=k-1\downarrow}^0 \hat{W} \, \exp(-\tau_q D) \right]$$

We define the conditional probability distribution (where $[\tau_q]_0^k \equiv [\tau_0, \dots \tau_k]$ denotes an ordered contiguous sequence of time intervals)

$$\Pr\left(I, [\tau_q]_0^k, k \mid J, t\right) = \left\{ \exp(-\tau_k D) \left[ \prod_{q=k-1\downarrow}^0 \hat{W} \, \exp(-\tau_q D) \right] \delta\left(t - \sum_{q=0}^k \tau_q\right) \right\}_{IJ} \tag{10}$$

For $D_{II} \neq 0$,

$$\Pr\left(I, [\tau_q]_0^k, k \mid J, t\right) = \left\{ \exp(-\tau_k D) \left[ \prod_{q=k-1\downarrow}^0 \left(\hat{W} \, D^{-1}\right) (D \exp(-\tau_q D)) \right] \delta\left(t - \sum_{q=0}^k \tau_q\right) \right\}_{IJ}$$

Either way,

$$\Pr(I \mid J, t) = \sum_{k=0}^{\infty} \int_0^t \cdots \int_0^t \left(\Pi_{q=0}^k \, d\tau_q\right) \Pr\left(I, [\tau_q]_0^k, k \mid J, t\right) = \left[\exp\left(t\left(\hat{W} - D\right)\right)\right]_{IJ}.$$

## 2.4 Semigroup property

Suppose $t = t_1 + t_2$, all nonnegative. Then for any time-evolution equation we must have the semigroup property:

$$\Pr(I \mid J, t) = \sum_K \Pr(I \mid K, t_2) \Pr(K \mid J, t_1)$$

i.e.

$$\left[\exp\left(t\left(\hat{W} - D\right)\right)\right]_{IJ} = \sum_K \left[\exp\left(t_2\left(\hat{W} - D\right)\right)\right]_{IK} \left[\exp\left(t_1\left(\hat{W} - D\right)\right)\right]_{KJ}.$$

Is there a $k$-event version of this rule, for $k = k_1 + k_2$? In other words, can we add (nonnegative) numbers of reaction events rather than time intervals?

We observe (where again $[\tau_q]_0^k \equiv [\tau_0, \ldots \tau_k]$)

$$\Pr(I, k \mid J, t) = \int_0^t \cdots \int_0^t (\Pi_{q=0}^k \, d\tau_q) \Pr(I, [\tau_q]_0^k, k \mid J, t)$$

and

$$\sum_K \int_0^t d\tau \, \Pr\big(I, [\tau'_{k_1}, \tau_{k_1+1}, \ldots \tau_k], k_2 \mid K, \tau\big) \Pr\big(K, [\tau_q]_0^{k_1}, k_1 \mid J, t - \tau\big)$$

$$= \int_0^t d\tau \left\{ \exp(-\tau_k D) \left[ \prod_{q=k-1\downarrow}^{k_1+1} \hat{W} \, \exp(-\tau_q D) \right] \left[ \hat{W} \, \exp(-\tau'_{k_1} D) \right] \right.$$

$$\times \delta\!\left( \tau - \left( \sum_{q=k_1+1}^{k} \tau_q + \tau'_{k_1} \right) \right) \exp(-\tau_{k_1} D) \left[ \prod_{q=k_1-1\downarrow}^{0} \hat{W} \, \exp(-\tau_q D) \right] \delta\!\left( t - \tau - \sum_{q=0}^{k_1-1} \tau_q \right) \right\}_{IJ}$$

$$= \left\{ \exp(-\tau_k D) \left[ \prod_{q=k-1\downarrow}^{k_1+1} \hat{W} \, \exp(-\tau_q D) \right] \left[ \hat{W} \, \exp\!\big( -(\tau'_{k_1} + \tau_{k_1}) D \big) \right] \right.$$

$$\times \left[ \prod_{q=k_1-1\downarrow}^{0} \hat{W} \, \exp(-\tau_q D) \right] \delta\!\left( t - \left( \sum_{q=k_1+1}^{k} \tau_q + \tau'_{k_1} + \sum_{q=0}^{k_1-1} \tau_q \right) \right) \right\}_{IJ}$$

$$= \Pr\big(I, [[\tau_q]_0^{k_1-1}, \tau'_{k_1} + \tau_{k_1}, [\tau_q]_{k_1+1}^k], k \mid J, t\big) .$$

Thus, if $k = k_1 + k_2$ and for any $\tau'_{k_1} \in [0, \tau_{k_1}]$, there is a semigroup law:

$$\Pr(I, [\tau_q]_0^k, k \mid J, t) =$$

$$\sum_K \int_0^t d\tau \, \Pr\big(I, [\tau'_{k_1}, \tau_{k_1+1}, \ldots \tau_k], k_2 \mid K, \tau\big) \Pr\big(K, [\tau_0, \ldots \tau_{k_1-1}, \tau_{k_1} - \tau'_{k_1}], k_1 \mid J, t - \tau\big) \tag{11}$$

In this derivation there is an arbitrary choice of $\tau'_{k_1}$ from the interval $[0, \tau_{k_1}]$. Consistent uniform possibilities include $\tau'_{k_1} = 0$ and $\tau'_{k_1} = \tau_{k_1}$. We will choose the latter option, so that the last $\tau$ in a series $[\tau_q]$ is zero. The result used in the next section will be a semigroup law for "just-fired" probabilities, right after a reaction or (more generally) a rule firing.

# 3    Results and discussion

Given the foregoing semigroup expression from TOPE, we complete the derivation of a Markov chain representing the SSA algorithm. We then consider extensions of this result, including parameterized reactants, but focussing mainly on hybrid stochastic event/ordinary differential equation dynamical systems.

## 3.1    Derivation of a Markov chain

### 3.1.1    Just-fired probabilities and Bayes' rule

Define the just-fired probabilities

$$\breve{\mathrm{Pr}}\big(I,\,[\tau_q]_0^{k-1},\,k\,|\,J,\,t\big) = \mathrm{Pr}\big(I,\,[\tau_0,\,\ldots\tau_{k-1},\,\tau_k=0],\,k\,|\,J,\,t\big)$$

$$= \left[\prod_{q=k-1\downarrow}^{0}\hat{W}\,\exp(-\tau_q\,D)\right]_{I\,J}\delta\!\left(t-\sum_{q=0}^{k}\tau_q\right)$$

and calculate their conditional distributions given the number of events $k$, using Bayes' rule :

$$\breve{\mathrm{Pr}}\big(I,\,[\tau_q]_0^{k-1}\,|\,J,\,t,k\big) = \mathrm{Pr}\big(I,\,[[\tau_q]_0^{k-1},\,\tau_k=0]\,\big|\,J,\,t,k\big)$$

$$= \frac{\mathrm{Pr}\big(I,\,[[\tau_q]_0^{k-1},\,0],\,k\,|\,J,\,t\big)}{\sum_{I'}\int_0^t\cdots\int_0^t\big(\Pi_{q=0}^{k-1}\,d\,\tau_q'\big)\,\mathrm{Pr}\big(I',\,[[\tau_q']_0^{k-1},\,0],\,k\,|\,J,\,t\big)}$$

$$= \frac{\left[\prod_{q=k-1\downarrow}^{0}\hat{W}\,\exp(-\tau_q\,D)\right]_{I\,J}\delta\big(t-\sum_{q=0}^{k}\tau_q\big)}{\sum_{I'}\int_0^t\cdots\int\big(\Pi_{q=0}^{k-1}\,d\,\tau_q'\big)\left[\prod_{q=k-1\downarrow}^{0}\hat{W}\,\exp(-\tau_q'\,D)\right]_{I'\,J}\delta\big(t-\sum_{q=0}^{k}\tau_q'\big)}\,.$$

With $\tau_{k_1}' = \tau_{k_1}$ and $\tau_k = 0$ the semigroup law becomes:

$$\mathrm{Pr}\big(I,\,[[\tau_q]_0^{k-1},\,0],\,k\,|\,J,\,t\big)$$

$$= \sum_K\int_0^t d\,\tau\,\mathrm{Pr}\big(I,\,[\tau_{k_1},\,\ldots\tau_{k-1},\,0],\,k_2\,|\,K,\,\tau\big)\,\mathrm{Pr}\big(K,\,[\tau_0,\,\ldots\tau_{k_1-1},\,0],\,k_1\,|\,J,\,t-\tau\big)$$

i.e.

$$\breve{\mathrm{Pr}}\big(I,\,[\tau_q]_0^{k-1},\,k\,|\,J,\,t\big) = \sum_K\int_0^t d\,\tau\,\breve{\mathrm{Pr}}\big(I,\,[\tau_q]_{k_1}^{k-1},\,k_2\,|\,K,\,\tau\big)\,\breve{\mathrm{Pr}}\big(K,\,[\tau_q]_0^{k_1-1},\,k_1\,|\,J,\,t-\tau\big) \qquad (12)$$

For $k_2 = 1$ and $k > 1$,

$$\breve{\mathrm{Pr}}\big(I,\,[\tau_q]_0^{k-1},\,k\,|\,J,\,t\big) = \sum_K\int_0^t d\,\tau\,\breve{\mathrm{Pr}}\big(I,\,\tau_{k-1},\,1\,|\,K,\,\tau\big)\,\breve{\mathrm{Pr}}\big(K,\,[\tau_q]_0^{k-2},\,k-1\,|\,J,\,t-\tau\big) \qquad (13)$$

Suppose now that we wish to sample from this dynamical system for a total time $t$ that is drawn from an exponential distribution with mean $T$:

$$\breve{\mathrm{Pr}}(t) = \exp(-t/T)/T$$

We will take $T$ to be much longer than other characteristic times in the system given by nonzero matrix elements of $W$. Then

$$\breve{\mathrm{Pr}}\big(I,\,[\tau_q]_0^{k-1},\,t,\,k\,|\,J\big) = \breve{\mathrm{Pr}}\big(I,\,[\tau_q]_0^{k-1},\,k\,|\,J,\,t\big)\,\breve{\mathrm{Pr}}(t) = T^{-1}\left[\prod_{q=k-1\downarrow}^{0}\hat{W}\,\exp(-\tau_q(D+\mathcal{I}/T))\right]_{I\,J}\delta\!\left(t-\sum_{q=0}^{k}\tau_q\right)$$

where $\mathcal{I}$ is the identity matrix.

Then

$$\breve{\mathrm{Pr}}\big(I,\,[\tau_q]_0^{k-1},\,k\,|\,J\big) = \int_0^\infty d\,t\,\breve{\mathrm{Pr}}\big(I,\,[\tau_q]_0^{k-1},\,t,\,k\,|\,J\big) = T^{-1}\left[\prod_{q=k-1\downarrow}^{0}\hat{W}\,\exp(-\tau_q(D+\mathcal{I}/T))\right]_{I\,J}$$

$$\breve{\mathrm{Pr}}\big(I,\,k\,|\,J\big) = T^{-1}\left[\prod_{q=k-1\downarrow}^{0}\hat{W}\,\int_0^\infty d\,\tau_q\exp(-\tau_q(D+\mathcal{I}/T))\right]_{I\,J} = T^{-1}\left[\prod_{q=k-1\downarrow}^{0}\hat{W}\,\,(D+\mathcal{I}/T)^{-1}\right]_{I\,J}$$

Assuming $T \gg k/D_{II}$ for every element $D_{II}$ of $D$ (assuming the reaction network has been modified as in [2] to formally eliminate terminal states),

$$(D + I/T)^{-1} = D^{-1}(I + D^{-1}/T)^{-1} \approx D^{-1}(I - D^{-1}/T + \ldots)$$

and

$$\tilde{\Pr}(I, k \mid J) = T^{-1}\left[(\hat{W} \ D^{-1})^k\right]_{I,J}(1 + O(k D^{-1}/T))$$

$$\tilde{\Pr}(k \mid J) = \sum_I \tilde{\Pr}(I, k \mid J) \approx T^{-1}(1 + O(k D^{-1}/T))$$

Small $t/T$:

$$\tilde{\Pr}(t \mid J) = \tilde{\Pr}(t) = \exp(-t/T)/T = T^{-1}(1 + O(t/T))$$

Bayes' rule:

$$\tilde{\Pr}\left(I, [\tau_q]_0^{k-1}, k \mid J, t\right) = \tilde{\Pr}\left(I, [\tau_q]_0^{k-1}, t \mid J, k\right) \frac{\tilde{\Pr}(k \mid J)}{\tilde{\Pr}(t)}$$

Substituting this expression into Equation 13 everywhere,

$$\tilde{\Pr}\left(I, [\tau_q]_0^{k-1}, t \mid J, k\right) \frac{\tilde{\Pr}(k \mid J)}{\tilde{\Pr}(t)}$$

$$= \sum_K \int_0^t d\tau \, \frac{\tilde{\Pr}(1 \mid K)}{\tilde{\Pr}(\tau)} \frac{\tilde{\Pr}(k-1 \mid J)}{\tilde{\Pr}(t-\tau)} \tilde{\Pr}(I, \tau_{k-1}, \tau \mid K, 1) \tilde{\Pr}\left(K, [\tau_q]_0^{k-2}, t-\tau \mid J, k-1\right)$$

$$\tilde{\Pr}\left(I, [\tau_q]_0^{k-1}, t \mid J, k\right) = \sum_K \int_0^t d\tau \left(\frac{\tilde{\Pr}(1 \mid K) \tilde{\Pr}(k-1 \mid J)}{\tilde{\Pr}(k \mid J)}\right)\left(\frac{\tilde{\Pr}(t)}{\tilde{\Pr}(\tau) \tilde{\Pr}(t-\tau)}\right)$$

$$\times \tilde{\Pr}(I, \tau_{k-1}, \tau \mid K, 1) \tilde{\Pr}\left(K, [\tau_q]_0^{k-2}, t-\tau \mid J, k-1\right)$$

But we can evaluate the ratios

$$\frac{\tilde{\Pr}(t)}{\tilde{\Pr}(\tau) \tilde{\Pr}(t-\tau)} = T$$

$$\frac{\tilde{\Pr}(1 \mid K) \tilde{\Pr}(k-1 \mid J)}{\tilde{\Pr}(k \mid J)} = T^{-1}(1 + O(t/T))$$

Thus to leading (zero'th) order in $t/T$,

$$\tilde{\Pr}\left(I, [\tau_q]_0^{k-1}, t \mid J, k\right) \approx \sum_K \int_0^t d\tau \, \tilde{\Pr}(I, \tau_{k-1}, \tau \mid K, 1) \tilde{\Pr}\left(K, [\tau_q]_0^{k-2}, t-\tau \mid J, k-1\right) \qquad (14)$$

and therefore (by integration over $\tau_q$)

$$\tilde{\Pr}(I, t \mid J, k) \approx \sum_K \int_0^t d\tau \, \tilde{\Pr}(I, \tau \mid K, 1) \tilde{\Pr}(K, t-\tau \mid J, k-1) \qquad (15)$$

This can be interpreted as an infinite-dimensional (first-order) Markov chain which increments $k$ each iteration, and which updates $I$ and $t$ according to the conditional probability operator

$$\mathcal{W}(I, t' \mid J, t) = \tilde{\mathrm{Pr}}(I, t' - t \mid J, 1) = \int_0^\infty d\tau' \, \tilde{\mathrm{Pr}}(I, \tau', t' - t \mid J, 1)$$

$$= \int_0^\infty d\tau' \, \tilde{\mathrm{Pr}}(I, \tau', 1 \mid J, t' - t) \, \frac{\tilde{\mathrm{Pr}}(t' - t)}{\tilde{\mathrm{Pr}}(1 \mid J)}$$

$$\approx \int_0^\infty d\tau' \, \tilde{\mathrm{Pr}}(I, \tau', 1 \mid J, t' - t) \, \frac{\exp(-(t' - t)/T)/T}{1/T} \ .$$

Using

$$\tilde{\mathrm{Pr}}(I, \tau', 1 \mid J, t' - t) = \left[ \hat{W} \, \exp(-\tau' D) \right]_{I,J} \delta(t' - t - \tau')$$

we calculate

$$\mathcal{W}(I, t' \mid J, t) \approx \int_0^\infty d\tau' \left[ \hat{W} \, \exp(-\tau' D) \right]_{I,J} \delta(t' - t - \tau') \exp(-(t' - t)/T)$$

$$= \left[ \hat{W} \, \exp(-(t' - t) D) \right]_{I,J} \exp(-(t' - t)/T) \, \Theta(t' \geq t)$$

$$= \left[ \hat{W} \, \exp(-(t' - t) (D + \mathcal{I}/T)) \right]_{I,J} \Theta(t' \geq t)$$

$$\approx \left[ \hat{W} \, \exp(-(t' - t) D) \right]_{I,J} \Theta(t' \geq t) \ .$$

Thus,

$$\mathcal{W}(I, t' \mid J, t) \approx \hat{W}_{I,J} \exp(-(t' - t) D_{J,J}) \, \Theta(t' \geq t) \ . \tag{16}$$

This expression is properly normalized:

$$\sum_I \int_{-\infty}^\infty d t' \, \mathcal{W}(I, t' \mid J, t) = D_{J,J} \int_{-\infty}^\infty d t' \exp(-(t' - t) D_{J,J}) \, \Theta(t' \geq t)$$

$$= D_{J,J} \int_0^\infty d\tau \exp(-\tau D_{J,J}) = 1$$

as claimed. For $D_{J,J} \neq 0$, we may factor $\mathcal{W}$ into a time update followed by a state update, each normalized:

$$\mathcal{W}(I, t' \mid J, t) \approx M_1 \, M_0 \quad \text{where}$$
$$M_1 \equiv \left( \hat{W}_{I,J} \, D_{J,J}^{-1} \right) \quad \text{and} \quad M_0 \equiv [D_{J,J} \exp(-(t' - t) D_{J,J}) \, \Theta(t' \geq t)] \ . \tag{17}$$

Here matrix $M_0$ is diagonal, so the elementwise product is also a matrix product. Thus

$$\tilde{\mathrm{Pr}}(I, t \mid J, k) \approx \sum_K \int_0^t d\tau \, \mathcal{W}(I, t \mid K, t - \tau) \, \tilde{\mathrm{Pr}}(K, t - \tau \mid J, k - 1)$$

$$= \sum_K \int_0^\infty d t_{\text{old}} \, \mathcal{W}(I, t \mid K, t_{\text{old}}) \, \tilde{\mathrm{Pr}}(K, t_{\text{old}} \mid J, k - 1), \quad \text{where} \tag{18}$$

$$\tilde{\mathrm{Pr}}(t \mid J, k) \equiv \mathcal{W} \circ \tilde{\mathrm{Pr}}(t - \tau \mid J, k - 1)$$

and finally the algorithmic Markov chain expression

$$\tilde{\mathrm{Pr}}(t \mid J, k) = \mathcal{W}^k \circ \tilde{\mathrm{Pr}}(0 \mid J, 0) \ . \tag{19}$$

### 3.1.2    SSA Theory Summary

We use the Time-Ordered Product Expansion with the diagonal decomposition:

$$W = \hat{W} - D \quad \text{and} \quad D = \mathrm{diag}\big(\mathbf{1} \cdot \hat{W}\big) \ .$$

Then an event-oriented simulation requires the use of Bayes' rule to convert from $P(a, [\tau_q \mid 0 \leq q \leq k - 1], k \mid c, t)$ to $P(a, [\tau_q \mid 0 \leq q \leq k - 1], t \mid c, k)$, and this in turn requires a distribution on total simulation time $t$. To this end, suppose there is a small constant probability per unit time, $\epsilon = 1/T_{\text{long}}$, that a simulation of the stochastic process defined by $W$ comes to an end. This termination process will provide an exponential prior distribution on the time variable $t$. Taking the limit in which $T_{\text{long}}$ is much longer than the time $t$ of interest to us, so that $\epsilon t = t/T_{\text{long}} \to 0$, the parameter $T_{\text{long}}$ drops out of the calculation and the Bayes-transformed simulation statistics are unaffected by the probability of pending termination.

Then an event-oriented simulation requires the use of Bayes' rule to convert from $P(a, [\tau_q \mid 0 \leq q \leq k-1], k \mid c, t)$ to $P(a, [\tau_q \mid 0 \leq q \leq k-1], t \mid c, k)$, and this in turn requires a distribution on total simulation time $t$. To this end, suppose there is a small constant probability per unit time, $\epsilon = 1/T_{\text{long}}$, that a simulation of the stochastic process defined by $W$ comes to an end. This termination process will provide an exponential prior distribution on the time variable $t$. Taking the limit in which $T_{\text{long}}$ is much longer than the time $t$ of interest to us, so that $\epsilon t = t/T_{\text{long}} \to 0$, the parameter $T_{\text{long}}$ drops out of the calculation and the Bayes-transformed simulation statistics are unaffected by the probability of pending termination.

Define a prior on the total simulation time with $T \to \infty$: $\tilde{\Pr}(t) = \exp(-t/T)/T$. Then from the TOPE, the just-fired probabilities form a Markov chain:

$$\tilde{\Pr}(I, k \mid J) = T^{-1} \left[ \prod_{q=k-1\downarrow}^{0} \hat{W} \int_0^\infty d\tau_q \exp(-\tau_q (D + \mathcal{I}/T)) \right]_{IJ}$$

$$= T^{-1} \left[ \prod_{q=k-1\downarrow}^{0} \hat{W} \ (D + \mathcal{I}/T)^{-1} \right]_{IJ}$$

$$\tilde{\Pr}(I, k \mid J) = T^{-1} \left[ \left( \hat{W} \ D^{-1} \right)^k \right]_{IJ} \left( 1 + O(k \, D^{-1}/T) \right)$$

The joint probabilities of system states and elapsed time also form a Markov chain:

$$\mathcal{W}(I, t' \mid J, t) \approx \left( \hat{W}_{IJ} \, D_{JJ}^{-1} \right) [D_{JJ} \exp(-(t'-t) D_{JJ}) \, \Theta(t' \geq t)] \quad \text{and}$$
$$\tilde{\Pr}(t \mid J, k) \equiv \mathcal{W} \circ \tilde{\Pr}(t - \tau \mid J, k-1) \, . \tag{20}$$

where the "$\circ$" inner product operation combines both a sum over all states and an integral over all nonnegative times. The off-diagonal operator is a sum of reaction propensities $k_J^{(r)}$ multiplied by reaction-specific matrices $S_{IJ}^{(r)}$ specifying their effects on particle numbers, so we may write

$$\hat{W}_{IJ} = \sum_r \hat{W}_{IJ}^{(r)} = \sum_r k_J^{(r)} \, S_{IJ}^{(r)}, \text{ where}$$

$$S_{IJ}^{(r)} \in \{0, 1\} \quad \text{and} \quad S_{IJ}^{(r)} = 1 \Rightarrow k_J^{(r)} > 0$$

Now using $S_{II}^{(r)} = 0$ and $\sum_I S_{IJ}^{(r)} \leq 1$, we have a Markov interpretation of $\left( \hat{W}_{IJ} \, D_{JJ}^{-1} \right)$ in Equation 20:

$$\hat{W}_{IJ} \, D_{JJ}^{-1} = \sum_r \text{Prob}(r \mid J) \, S_{IJ}^{(r)} = \text{Prob}(I \mid J) \, , \text{ where}$$

$$\text{Prob}(r \mid J) = \frac{k_J^{(r)}}{k_J^{(\text{total})}} \quad \text{and} \quad k_J^{(\text{total})} = \sum_{r'} k_J^{(r')} \, .$$

Of course the factor $[D_{JJ} \exp(-(t'-t) D_{JJ}) \, \Theta(t' \geq t)]$ in Equation 20 is just the SSA exponential distribution of non-negative waiting times between reaction events. Finally iterating the recursion relation of Equation 20, we find the algorithmic Markov chain expression for SSA (Equation 19) : $\tilde{\Pr}(t \mid J, k) = \mathcal{W}^k \circ \tilde{\Pr}(0 \mid J, 0)$, which expresses the iteration of a Markov chain.

This Markov chain expression has also been used as the starting point for the derivation of *exact* accelerated stochastic simulation algorithms [10,11] that execute many reactions per step (i.e. they "leap" forward) and thus go much faster than SSA, while also sampling from the exact probability distribution given by the just-fired probabilities above. These derivations proceed by algebraic rearrangement of terms to express computationally efficient versions of rejection sampling. The algorithm of [11] has been parallelized, which is often difficult for discrete-event simulations.

The foregoing derivation was outlined in far less detail in [8]. A similar equation for SSA was reached by very different methods in [9], Theorem 10.1. To our knowledge this is the first complete derivation of SSA from field theory methods like TOPE.

## 3.2     Algorithm: SSA

The SSA algorithm represented by the Markov chain in Equations 19 and 20 above may be written out in pseudocode as follows:

**repeat** {

    *compute* propensities $k^{(r)}$

    *compute* $k^{(\text{total})} = \sum_r k^{(r)}$

    **draw** waiting time $\Delta t$ from $k^{(\text{total})} \exp(-\Delta t\, k^{(\text{total})})$

    $t := t + \Delta t;$  // *advance* the clock by $\Delta t$

    **draw** reaction $r$ from distribution $k^{(r)} / k^{(\text{total})}$ and *execute* reaction $r$

} **until** $t \geq t_{\max}$

## 3.3     Extension: Parameterized rule and graph grammar SSA-like algorithm

For biological modeling, including spatial and mechanical modeling of biological systems, it is very important to generalize from pure particles to particles with both discrete and continuous attributes. The complication is that reaction or process rates can then depend on the attributes both of the incoming and outgoing objects. A non-molecular example of such parameterized objects would be cells of a given size, whose propensity to divide may actually depend on their real-valued size parameter (as in Section 3.5.5). More generally, this capability enables agent-based modeling and simulation since it allows interacting objects to have dynamic internal state and even (as explained in Section 3.3.2 below) dynamic relationships.

Generalizing from Equation 7, as in [4], we include all possible instantiations of parameters $x$ and $y$, the integrated off-diagonal process representation operator

$$\hat{O}_r = \int \int d\{x\}\, d\{y\}\, \rho_r([x_p], [y_q]) \left[ \prod_{q \in \text{rhs}(r)} \hat{a}_{\beta(q)}(y_q) \right] \left[ \prod_{p \in \text{lhs}(r)} a_{\alpha(p)}(x_q) \right] \tag{21}$$

The generalization is conceptually trivial because we have simply used a function $\rho_r([x_a], [y_b])$ to express the possibly infinite number of different reaction rates that pertain to objects that differ only in their attributes. Again (as in Equation 7 ), summation over all discrete-valued parameters and integration over all continuous-valued parameters generalizes the operator to handle all possible sets of parameter values.

### 3.3.1     Algorithm: SSA with parametrized reactant objects

The resulting variant of the SSA algorithm for parameterized reactions can be expressed in pseudocode as follows (outlined briefly in [8]):

**forall**  reactions $r$ *factor* $\rho^{(r)}(x_{\text{in}}, x_{\text{out}}) = k^{(r)}(x_{\text{in}})\, p^{(r)}(x_{\text{out}} \mid x_{\text{in}});$

**repeat** {

    *compute* SSA propensities as $k^{(r)}(x_{\text{in}});$

    *compute* $k^{(\text{total})} = \sum_r k^{(r)}(x_{\text{in}});$

    **draw** waiting time $\Delta t$ from $k^{(\text{total})} \exp(-\Delta t\, k^{(\text{total})})$ ;

    $t := t + \Delta t;$  // *advance* the clock by $\Delta t$

    **draw** reaction $r$ from distribution $k^{(r)}(x_{\text{in}}) / k^{(\text{total})}$ ;

    **draw** $x_{\text{out}}$ from $p^{(r)}(x_{\text{out}} \mid x_{\text{in}})$ and execute reaction $r$;

} **until** $t \geqslant t_{max}$

### 3.3.2    Structural matching

The functions $\rho(x, y)$ appearing in Equation 21 may impose constraints including equality of variables; equivalently we may allow some variables to appear multiple times in object parameter lists. Either way there follows a mechanism to encode structural relations - graphs and labelled graphs - in the input and output variable lists. Object attributes can include Object ID codes which other objects can also include in their parameter lists. (Of course, the numeric values of Object IDs can be *globally* permuted without changing the structural relationships among extant objects.) In this way, the integrated version of the parameterized reaction operator above encodes structural pattern matching, including variable-binding in logical formulate, among the preconditions that can be enforced before such a generalized reaction or "rule" can fire.

From this point of view, syntactic *variable-binding has the semantics of multiple integration* [4]. In this way we can entrain pattern-matching systems such as the computer algebra system *Mathematica*, or logic-based programming languages, to the job of simulating complex process rules. As in rule-based expert systems, when multiple rules might fire the Rete algorithm [10] can be used to speed up the computations required to maintain knowledge of their relative rates.

The resulting systems have the power to model and simulate dynamic labelled graphs including growing multicellular tissues with dynamical cell-neighbor relationships [4] and molecular complexes with dynamical binding structure [12,13,14]. Thus, the TOPE operator algebra approach also explains why and how structural (graph-) matching computations arise naturally in biochemical and multicellular biological simulation.

## 3.4    Hybrid SSA/ODE setup

As will be shown in Section 3.4 below, the operator formulation for a system of ordinary differential equations is [4]:

$$\hat{O}_{drift} = - \int \int d\{x\} \, d\{y\} \, \hat{a}(\{y\}) \, a(\{x\}) \left[ \sum_i \nabla_{y_i} \left( v_i(\{y\}) \prod_k \delta(y_k - x_k) \right) \right] \tag{22}$$

Here and in the calculations that follow, the Dirac delta function can be considered as a Gaussian with very small variance, which participates in a limiting process by which, at the end of each calculation, the limit of zero variance is taken.

In [4] this operator expression is generalized from ordinary differential equations to stochastic differential equations, for example those pertaining to the diffusion of particles, as equivalently represented by the Fokker-Planck equation.

### 3.4.1    Computation of matrix elements

From the commutator

$$[a(y), \, \hat{a}(x)] = \delta(y - x) \left( I + Q(N(x) \mid n_{max}) \, N(x) \right),$$

we may calculate matrix elements of $\hat{O}_{drift}$ in Equation 22 such as:

$$\langle w \mid \hat{O}_{\text{drift}} \mid z \rangle = -\left\langle \{w\} \middle| \int d\{x\} \int d\{y\} \left( \sum_i \nabla_{y_i} v_i(\{y\}) \, \delta(\{y\} - \{x\}) \right) \hat{a}(\{y\}) \, a(\{x\}) \, \hat{a}(\{z\}) \middle| 0 \right\rangle$$

$$= -\left\langle \{w\} \middle| \int d\{x\} \int d\{y\} \left( \sum_i \nabla_{y_i} v_i(\{y\}) \, \delta(\{y\} - \{x\}) \right) \hat{a}(\{y\}) \, \delta(\{x\} - \{z\}) [I + Q(N(x)) \, N(\{x\})] \middle| 0 \right\rangle$$

$$= -\int d\{x\} \int d\{y\} \left( \sum_i \nabla_{y_i} v_i(\{y\}) \, \delta(\{y\} - \{x\}) \right) \delta(\{x\} - \{z\}) \, \langle \{w\} \mid \{y\} \rangle$$

$$= -\int d\{y\} \left( \sum_i \nabla_{y_i} v_i(\{y\}) \, \delta(\{y\} - \{z\}) \right) \delta(\{w\} - \{y\})$$

$$= +\int d\{y\} \, \delta(\{y\} - \{z\}) \left( \sum_i v_i(\{y\}) \, \nabla_{y_i} \, \delta(\{w\} - \{y\}) \right) - \int_{\partial} d\{y\} \left( \sum_i v_i(\{y\}) \, \delta(\{y\} - \{z\}) \, \delta(\{y\} - \{w\}) \right)$$

$$= \sum_i v_i(\{z\}) \, \nabla_{z_i} \, \delta(\{w\} - \{z\}) + \text{boundary term} \; (\rightarrow 0 \; \text{here})$$

The easiest treatment for the boundary terms is to add the assumptions that boundaries are at infinity in the space of parameters $x$, $y$ and $z$, and that initial conditions place zero probability there, and that finite velocities $v(x)$ ensure the probability remains zero at infinity at finite times. In that case boundary terms can be neglected. Alternatively, we can define $O_{\text{drift}} = \hat{O}_{\text{drift}} - \text{diag}(1 \cdot \hat{O}_{\text{drift}})$ which in this case subtracts off the boundary term. Then

$$O_{\text{drift}} = -\int \int d\{x\} \, d\{y\} \, (\hat{a}(\{y\}) \, a(\{x\}) - \hat{a}(\{x\}) \, a(\{x\})) \left[ \sum_i \nabla_{y_i} \left( v_i(\{y\}) \prod_k \delta(y_k - x_k) \right) \right] \tag{23}$$

If we define $x(t)$ as a time-varying version of $z$, satisfying

$$\frac{\partial x_i}{\partial t} = v_i(\{x_k\}),$$

then

$$\sum_i v_i(\{z\}) \, \nabla_{x_i} \, \delta(\{w\} - \{x(t)\}) = \sum_i \left( \frac{\partial x_i}{\partial t} \right) \left( \frac{\partial}{\partial x_i} \right) \delta(\{w\} - \{x(t)\}) = \left( \frac{d}{dt} \right) \delta(\{w\} - \{x(t)\})$$

Next we calculate $\langle w \mid \exp(\tau \, O_{\text{drift}}) \mid z \rangle$. To this end, Taylor's theorem may be written

$$\text{Shift}_\tau \circ f(t) = f(t + \tau) \simeq \sum_{n=0}^{\infty} \frac{\tau^n}{n!} \left( \frac{d}{dt} \right)^n f(t) \simeq e^{(\tau D_i)} f(t)$$

if $\tau$ is a constant. For small $\tau$ we have

$$\langle w \mid \exp(\tau \, O_{\text{drift}}) \mid x \rangle = \langle w \mid x \rangle + \tau \, \langle w \mid O_{\text{drift}} \mid x \rangle + O(\tau^2)$$

$$= \left( 1 + \tau \left( \frac{d}{dt} \right) \right) \delta(\{w\} - \{x(t)\}) + O(\tau^2)$$

$$= \text{Shift}_\tau \, \delta(\{w\} - \{x(t)\}) \equiv \delta(\{w\} - \{x(t + \tau)\}) + O(\tau^2)$$

For larger $\tau$ we have

$$\langle w \mid \exp(\tau \, O_{\text{drift}}) \mid x \rangle = \lim_{n \to \infty} \left\langle w \middle| \prod_{i=1}^{n} \exp\left( \frac{\tau}{n} \, O_{\text{drift}} \right) \middle| x \right\rangle$$

$$= \int \cdots \int dx_{n-1} \cdots dx_1 \, [\text{Shift}_{\tau/n} \, \delta(\{w\} - \{x_{n-1}\})] \cdots [\text{Shift}_{\tau/n} \, \delta(\{x_1\} - \{x\})]$$

$$= \lim_{n \to \infty} \left( \prod_{i=1}^{n} \mathrm{Shift}_{\tau/n} \right) \delta(\{w\} - \{x(t)\})$$

$$= \mathrm{Shift}_\tau \, \delta(\{w\} - \{x(t)\}) \equiv \delta(\{w\} - \{x(t+\tau)\})$$

$$= \delta\left( \{w\} - \left( z(t=0) + \int_0^t v_i \, (z(t)) \, dt \right) \right)$$

Thus (where "IC" means initial condition)

$$\langle w \mid \exp(t \, O_{[DE]}) \mid z \rangle = \exp\left( t \sum_i v_i(\{z\}) \, \nabla_{z_i} \right) \delta(\{w\} - \{z\})$$

$$= \delta\left( \{w\} - \left( \{z(0) = z\} + \int_0^t v_i \, (z(t')) \, dt \right) \right) \tag{24}$$

$$= \delta\left( \{w\} - \left( \text{Solution of } \frac{\partial \, x_i}{\partial \, t} = v_i(\{x_k\}) \text{ with IC } z(0) = z \right) \right)$$

QED.

As far as we know this derivation has not appeared previously. As a corrollary, using Equation 24 we may multiply by $f(w)$ and integrate over $w$ to calculate

$$\exp(t \, v \, (\{z\}) \cdot \nabla_z) \, \delta \, (w - z) = \delta\left( w - \left( z(0) + \int_0^t v(z(t')) \, dt' \right) \right)$$

$$\Rightarrow \quad \exp(t \, v \, (\{z\}) \cdot \nabla_z) \, f \, (z) = f\left( z(0) + \int_0^t v(z(t')) \, dt' \right). \tag{25}$$

## 3.5 Hybrid SSA/ODE: Operator algebra derivation

We now derive a new SSA-like simulation algorithm for hybrid combinations of discrete events and ODE dynamics, using operator algebra. The main idea is to replace the exponential distribution factor $\exp(-t \, D)$ with $\exp\left( -\int_0^t D(t') \, dt' \right)$, and add an extra ODE to the system of ODEs in order to keep track of the integral. We will use the more compact formulation of TOPE in Equation 2 to derive this method.

### 3.5.1 Heisenberg picture

Let the operators, rather than the states, evolve in time according to $W_0$ according to Equation 3. This is traditionally called the "Heisenberg picture" in distinction to the "Schroedinger picture" in quantum mechanics. Recall Equation 2 and Equation 3, where $(\cdots)_+$ is the time-ordering super-operator :

$$(O(t_i) \, O \, (t_j) \, )_+ = \begin{cases} O(t_i) \, O \, (t_j) & \text{if } \; t_i \geqslant t_j \\ O(t_j) \, O \, (t_i) & \text{if } \; t_i \leqslant t_j \end{cases}$$

(and likewise for higher order products). Note that if $O(t_i)$ and $O(t_j)$ commute for all pairs of times $t_i$ and $t_j$, then $(O(t_i) \, O \, (t_j) \, )_+ = O(t_i) \, O \, (t_j)$, and the time-ordering operator $(\cdots)_+$ can be dropped.

### 3.5.2 Application to ODE + decay clock

The hybrid system consisting of chemical reactions (possibly parameterized) together with ordinary differential equations has the combined operator $W = \left( \hat{O}_{\text{react}} - D_{\text{react}} \right) + O_{\text{DE}}$, which we can regroup as

$$W = (O_{\text{DE}} - D_{\text{react}}) + \hat{O}_{\text{react}}$$

and then apply TOPE to $O_{\mathrm{DE}} - D_{\mathrm{react}}$ first with $W_{00} = O_{\mathrm{DE}}$ and $W_{01}(t_k) = -D_{\mathrm{react}}(t_k)$, and then again to $(O_{\mathrm{DE}} - D_{\mathrm{react}}) + \hat{O}_{\mathrm{react}}$ with $W_0 = W_{00} + W_{01} = O_{\mathrm{DE}} - D_{\mathrm{react}}$ and $W_1 = \hat{O}_{\mathrm{react}}$.

In the first application of TOPE to $O_{\mathrm{DE}} - D_{\mathrm{react}}$ with $W_{00} = O_{\mathrm{DE}}$, the opererators $W_{01}(t_k) = -D_{\mathrm{react}}(t_k)$ defined at different times are all diagonal in the same (particle number basis and therefore commute:

$$[D_{\mathrm{react}}(t_i), D_{\mathrm{react}}(t_j)] = 0 .$$

In this circumstance, we can *simply drop* the time-ordering super-operator $(\cdots)_+$ in Equation 2 and write

$$\exp(t\,(O_{\mathrm{DE}} - D_{\mathrm{react}})) = \exp(t\,O_{\mathrm{DE}}) \exp\left(-\int_0^t dt'\, D_{\mathrm{react}}(t')\right) \tag{26}$$

where, as in Equation 3, $D_{\mathrm{react}}(t') = \exp(-t'\,O_{\mathrm{DE}})\,D_{\mathrm{react}}\exp(t'\,O_{\mathrm{DE}})$ .In our case, (Equation 26) specializes to :

$$\langle w\,|\exp(t\,(O_{[\mathrm{DE}]} - D_{\mathrm{react}}))\,|\,z\rangle = \exp\left(t\sum_i v_i(\{z\})\,\nabla_{z_i}\right)\exp\left(-\int_0^t dt'\, D_{\mathrm{react}}(t')\right)\delta(\{w\} - \{z\})$$

$$= \exp\left(-\int_0^t dt'\, D_{\mathrm{react}}\left(z(0) + \int_0^{t'} v(\{z\})\,dt''\right)\right)\delta\left(w - \left(z(0) + \int_0^t v(z(t'))\,dt'\right)\right) \tag{27}$$

This result looks very similar to Equation 25 applied to $f(z) = \exp\left(-\int_0^t dt'\, D_{\mathrm{react}}(t')\right)\delta(\{w\} - \{z\})$, and we now aim to understand and exploit this similarity.

### 3.5.3 Equivalent ODE

Consider the dynamics expressed in Equation 27. Can we obtain the first factor from ODE's alone? Yes, if we introduce a new state variable $\tau$ involved in every ODE-related rule. Set $\tau(0) = 0$ as the new variable's initial condition, and augment the ODE operators as follows

$$\begin{aligned}
Z &= (z, \tau)\\
V(z) &= (v(\{z\}), -D(z))\\
\nabla_Z &= (\nabla_z, \partial_\tau)\\
\tilde{O}_{[\mathrm{DE}]} &= V(Z)\,\nabla_Z = v(\{z\})\cdot\nabla_z + D(z)\,\partial_\tau
\end{aligned} \tag{28}$$

In other words, we have added a differential equation for $\tau$ to the ODE system

$$\frac{\partial x_i}{\partial t} = v_i(\{x_k\}) \quad\text{and}\quad \frac{d\tau}{dt} = D(z) . \tag{29}$$

This equation is solvable in terms of a "warped time" coordinate

$$\tau(t) = \int_0^t D_{\mathrm{react}}(z(t'))\,dt' .$$

There are degenerate cases $D_{\mathrm{react}} = 0$ only if there are terminal states in the reaction network.

To see that this is the correct procedure, calculate from Equation 24:

$$\left\langle \left( \begin{matrix} w \\ \tau_{max} \end{matrix} \right) \middle| \exp\left(t\, \tilde{O}_{\{DE\}}\right) \exp(-\tau_{max}) \middle| \left( \begin{matrix} z(0) \\ \tau(0)=0 \end{matrix} \right) \right\rangle$$

$$= \exp\left( t\left( \sum_i v_i(\{z\})\, \nabla_{z_i} + D(z)\, \partial_\tau \right)\right) \delta(\{w\}-z)\, \delta(\tau_{max}-\tau)\, \exp(-\tau_{max})$$

$$(30)$$

$$= \delta\left(\{w\} - \left(z(0) + \int_0^t v_i\left(z(t')\right) dt'\right)\right)\delta\left(\tau_{max} - \int_0^t D_{react}(z(t'))\, dt'\right) \exp(-\tau_{max})$$

$$= \delta\big(\{w\} - \big(\text{Solution of Equation 29 , with IC } z(0),\ \tau(0)=0\big)\big) \times \exp(-\tau_{max})$$

This expression agrees with Equation 27, as required. But how do we insure the IC on $\tau$? That can be done as follows:

$$\left\langle \left( \begin{matrix} w \\ \tau_{max} \end{matrix} \right) \middle| \exp\left(t\, \tilde{O}_{\{DE\}}\right) \exp(-\tau_{max}) \middle| \left( \begin{matrix} z \\ 0 \end{matrix} \right) \right\rangle =$$

$$\left\langle \left( \begin{matrix} w \\ \tau_{max} \end{matrix} \right) \middle| \exp\left(t\, \tilde{O}_{\{DE\}}\right) \exp(-\tau_{max}) \middle| \left( \begin{matrix} z \\ 0 \end{matrix} \right) \right\rangle \times \left( 1 = \int d\tau'\, d z' \left\langle \left( \begin{matrix} z' \\ \tau' \end{matrix} \right) \middle| \left( \begin{matrix} z \\ \tau \end{matrix} \right) \right\rangle \right)$$

$$= \left\langle \left( \begin{matrix} w \\ \tau_{max} \end{matrix} \right) \middle| \exp\left(t\, \tilde{O}_{\{DE\}}\right) \exp(-\tau_{max}) \left( \int d\tau'\, d z' \middle| \left( \begin{matrix} z' \\ 0 \end{matrix} \right) \right\rangle \left\langle \left( \begin{matrix} z' \\ \tau' \end{matrix} \right) \middle| \right) \middle| \left( \begin{matrix} z \\ \tau \end{matrix} \right) \right\rangle$$

$$(31)$$

$$= \left\langle \left( \begin{matrix} w \\ \tau_{max} \end{matrix} \right) \middle| \exp\left(t\, \tilde{O}_{\{DE\}}\right) \exp(-\tau_{max})\, P_{\tau:=0} \middle| \left( \begin{matrix} z \\ \tau \end{matrix} \right) \right\rangle$$

where

$$P_{\tau:=0} \equiv \int d\tau'\, d z' \middle| \left( \begin{matrix} z' \\ 0 \end{matrix} \right) \right\rangle \left\langle \left( \begin{matrix} z' \\ \tau' \end{matrix} \right) \middle|$$

is a projection operator (i.e. one that satisfies $P \cdot P = P$) that resets the variable $\tau$ to zero after each use. In summary,

$$\left\langle \left( \begin{matrix} w \\ \tau_{max} \end{matrix} \right) \middle| \exp\left(t\, \tilde{O}_{\{DE\}}\right) \exp(-\tau_{max})\, P_{\tau:=0} \middle| \left( \begin{matrix} z \\ \tau \end{matrix} \right) \right\rangle =$$

$$\delta\big(\{w\} - \big(\text{Solution of Equation 29 , with IC } z(0),\ \tau(0)=0\big)\big) \times \exp(-\tau_{max})$$

$$(32)$$

Clearly this result is equivalent to Equation 27 and is in the correct form for a Markov chain that can represent a computation. Of course, the matrix element calculated is only relevant if $\tau_{max}$ as drawn from the exponential is constrained to be equal to the final value of $\tau$ in the final state $\left\langle \left( \begin{matrix} w \\ \tau_{max} \end{matrix} \right) \middle| \right.$ as solved by the ODE system $\tilde{O}_{\{DE\}}$. We can implement this constraint with a factor of $\delta(\tau - \tau_{max})$ in the Markov chain over states and times. Thus a step in the Markov chain between reactions can be written as:

$$\mathcal{W}_{between\ reactions} = \delta(\tau - \tau_{max}) \exp\left(t\, \tilde{O}_{\{DE\}}\right) P_{\tau:=0} \exp(-\tau_{max})\, \Theta(\tau_{max} \geqslant 0)$$

$$\mathcal{W} = \hat{O}_{react} \cdot \mathcal{W}_{between\ reactions}$$

As in the SSA derivation, the reaction step is given by factors of $\hat{O}_{react}$ which need to be normalized by $D_{react}$. Using $\delta(t - t_{max}(\tau_{max}))\, dt = \delta(\tau - \tau_{max})\, d\tau$ and $d\tau/dt = D_{react}(t)$, we find

$$M_{01} = \exp(-\tau_{max})\, \Theta(\tau_{max} \geqslant 0)$$

$$M_{00} = \delta(t - t_{max}(\tau_{max})) \exp\left(t\, \tilde{O}_{\{DE\}}\right) \cdot P_{\tau:=0}$$

$$M_1 = \hat{O}_{react} / D_{react}$$

$$\mathcal{W} = M_1 \cdot M_{00} \cdot M_{01}$$

$$(33)$$

where $\mathcal{W}$ represents the Markov chain corresponding to the simulation algorithm.

In implementations so far [14] we have used instead the equivalent differential operator

$$\tilde{O}_{\{DE\}} = V(Z)\, \nabla_Z = v(\{z\}) \cdot \nabla_z - D(z)\, p\, \partial_p$$

with $p = \exp(-\tau)$, initialized at $p_0 = 1$, and a uniform distribution on $p_{final} \in [0, 1]$.

### 3.5.4 Algorithm: Hybrid SSA/ODE solver

By Equation 33 above, a Markov chain algorithm for simulating the hybrid system can be represented in the following SSA-like pseudocode:

*factor* $\rho^{(r)}(x_{\text{in}}, x_{\text{out}}) = k^{(r)}(x_{\text{in}}) p^{(r)}(x_{\text{out}} \mid x_{\text{in}})$;

**repeat** {

    *initialize* SSA propensities as $k^{(r)}(x_{\text{in}})$;

    *initialize* $k^{(\text{total})} := \sum_r k^{(r)}(x_{\text{in}})$;

    *initialize* $\tau := 0$ ;

    **draw** effective waiting time $\tau_{\max}$ from $\exp(-\tau_{\max})$

    **solve** ODE system, including an extra ODE updating $\tau$:

        $\frac{d\tau}{dt} = k^{(\text{total})}(t)$

        **until** $\tau = \tau_{\max}$

    **draw** reaction $r$ from distribution $k^{(r)}(x_{\text{in}}) / k^{(\text{total})}$ ;

    **draw** $x_{\text{out}}$ from $p^{(r)}(x_{\text{out}} \mid x_{\text{in}})$ and *execute* reaction $r$;

} **until** $t \geq t_{\max}$

### 3.5.5 Application: Cell division

As a simplified model of stochastic cell division, we may consider constant growth of a linear dimension $l$ of each cell, $d\,l/d\,t = v$, coupled with a stochastic cell division rule whose propensity depends on the ratio of $l$ to a threshold length $l_0$ for likely division:

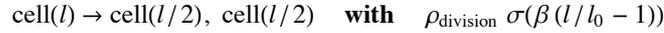

$$\text{cell}(l) \to \text{cell}(l/2),\ \text{cell}(l/2) \quad \textbf{with} \quad \rho_{\text{division}}\,\sigma(\beta\,(l/l_0 - 1))$$

with a sigmoidal function such as $\sigma(x) = 1/(1 + \exp(-x))$. In this model the parameter $\beta$ varies the sharpness of the threshold, and $\rho_{\text{division}}$ is the maximal propensity for division. Experimental evidence for stochastic dependence of division events on cell size in plant cells is reviewed in [15].

The differential equation for length can also be put in the form of a reaction rule that includes an ODE:

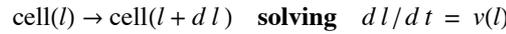

$$\text{cell}(l) \to \text{cell}(l + d\,l) \quad \textbf{solving} \quad d\,l/d\,t = v(l)$$

as described in [14]. Clearly this model could be augmented with other parameters such as growth signals with their own dynamics. This was done in models of biological stem cell niches in mouse olfactory epithelium and plant root growth, using the foregoing cell division rules. These systems were studied and simulated using the hybrid SSA/ODE algorithm above, in [14,16].

### 3.5.6 Application: Time-varying propensity for complete polymerization

Consider the n-step polymerization reaction

$$\{A \to X_1 \ \textbf{with}\ k_1,\ X_1 \to X_2 \ \textbf{with}\ k_2,\ \ldots,\ X_{n-1} \to B \ \textbf{with}\ k_n\}$$

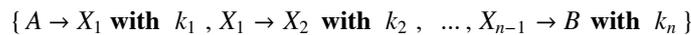

$$\tau_i = \tau/n \quad \text{and} \quad k_i = n\,k.$$

There is an $n^{(\max)}$ .

$$\hat{W} = \lambda\,\hat{a}_{n+1}$$

$$W = \lambda((\hat{a}_{n+1} + c_{n+1}) - I_{n+1}) = \lambda(\hat{b}_{n+1} - I_{n+1})$$

where $c_{n+1}$ is all zeros except for a "1" entry in the lower right corner. Since $\hat{b}$ and $I$ are matrices that commute, $\exp(t\,W) = \exp(t\,\lambda\,\hat{b}_{n+1})\exp(-t\,\lambda\,I_{n+1})$ and we easily compute

$$P(t\mid\tau, n) = [\exp(t\,W)]_{1,n+1} = \frac{\lambda^n\,t^{n-1}\,e^{-\lambda\,t}}{(n-1)!}$$

This is the distribution on polymer completion times. It is an Erlang distribution (a Gamma distribution with integral values of $n$). If $\tau$ is held fixed and $n$ tends towards infinity, this distribution approaches a delta function $\delta(t-\tau)$, which can lead to differential-delay equation models for reaction networks involving polymerization processes such as transcription [17]. This probability distribution for termination times also corresponds to the time-varying propensity function

$$\rho(t\mid\tau, n) = P(t\mid\tau, n) \Big/ \left[1 - \int_0^t P(t\mid\tau, n)\,dt\right] = \frac{\lambda^n\,t^{n-1}\,e^{-\lambda\,t}}{\Gamma(n, t\,\lambda)}\ , \qquad (34)$$

which increases monotonically in time.

As in Equation 18, the resulting time-varying propensity still fits within the framework of a Markov chain $\mathcal{W}(I, t'\mid J, t)$ that advances the time variable by an increment that is a random variable. The method of the previous section can be used to implement an SSA-like algorithm, with differential equations that govern propensities replaced by algebraic equations (Equation 34) or, if differential equations are also present, by differential-algebraic equations.

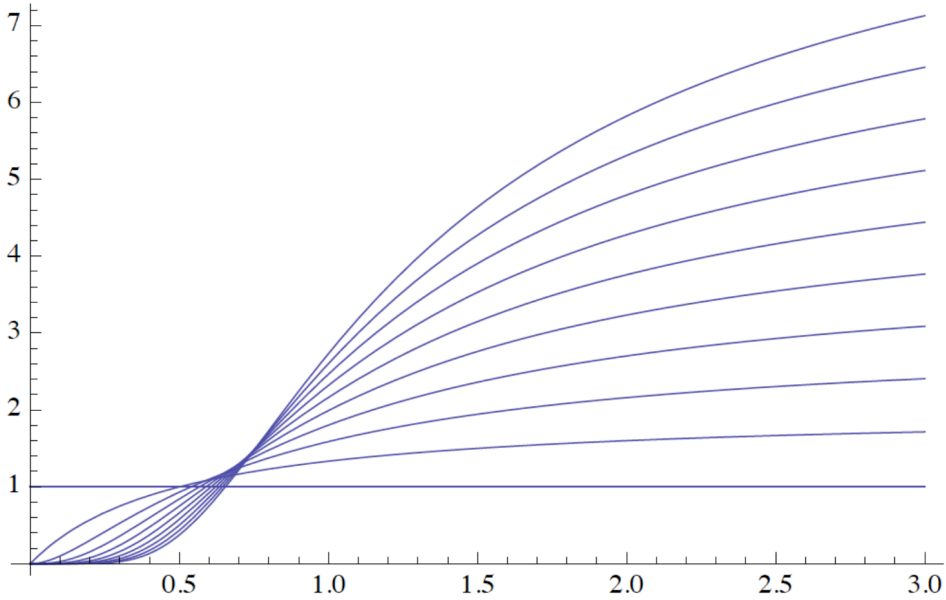

Figure 2. Erlang-derived time-dependent propensities for completion of a multistage process $\tau = 1$, $n \in \{1, \ldots, 10\}$. Horizontal axis: time, $t$. Vertical axis: propensity, $\rho(t\mid\tau, n)$. Plots for varying $n$ are superimposed. For larger $n$ there is a "maturation" phenomenon whereby completion at small times is very unlikely, and when a process is "overdue" for completion then its propensity becomes very high. By comparison, propensities for very small $n$ increase rapidly at first and are then relatively flat.

# 4    Conclusion and outlook

We have shown that the time-ordered product expansion (TOPE) can be used systematically to derive computational simulation and parameter-fitting algorithms for stochastic systems, connecting two seemingly distant areas of research. In doing so we have developed the means to translate formally between field theory language and the language of computable Markov chains in which randomized algorithms can be expressed and derived. By this means we hope to open the door to the use of TOPE and related methods from quantum and statistical field theory in the computational simulation of stochastic biochemical kinetics, with broad applicability in physically based biology. The particular hybrid stochastic process/ordinary differential equation simulation algorithm derived here is very different from interleaving and operator splitting algorithms which are intrinsically approximate; instead, this algorithm is exact in the same sense that SSA is (that is, it draws from the same distribution of just-fired reactions), except for errors introduced by the ODE solver and in the solver's detection of the ODE stopping criterion, which is that some variable reaches a threshold value. A future prospect for the field theory approach is for application to reaction-diffusion systems in which the propagator for particles between reactions is the heat kernal Green's function for the diffusion equation. The result may be an alternative avenue for derivation of novel particle-based, off-grid stochastic numerical solvers for reaction-diffusion problems as treated in [2], which, like the algorithms shown here, are also amenable to generalizations to exact "leaping" acceleration and to hybrid stochastic/differential equation solution algorithms.

# 5    Acknowledgements

Research was supported by NIH grants R01 GM086883 and P50 GM76516 to UC Irvine. I also wish to acknowledge the hospitality, travel support, and research environments provided by the Center for Nonlinear Studies (CNLS) at the Los Alamos National Laboratory, the Sainsbury Laboratory Cambridge University, and the Pauli Center for Theoretical Studies at ETH Zürich and the University of Zürich.

# A    References

# 6     Appendix: Maximum likelihood parameter inference

Application of the TOPE to maximum-likelihood parameter learning in stochastic reaction networks has previously been presented [16]. Here, for completeness of presentation for a different audience, we just show the essential gradient calculation step.

Suppose we have observations of the state of a chemical reaction network at times $\{t_s\}$, and wish to improve the probability $P(\text{Data} \mid \text{Model})$ of a reaction network model for the flow of probability at intermediate times. We will use the TOPE for each time interval in between observation times $t_s$ :

$$
\left[ e^{(t_{s+1} - t_s)\hat{W}} \right] (x(t_{s+1}), x(t_s)) = \sum_{k=0}^{\infty} \int_{t_s}^{t_{s+1}} \cdots \int_{t_s}^{t_{s+1}} d[\tau]_0^n \, \delta\!\left( (t_{s+1} - t_s) - \sum_{p=0}^{n} \tau_p \right)
$$
$$
\times \left[ e^{\tau_n \hat{D}} \, \hat{W} \dots e^{\tau_1 \hat{D}} \, \hat{W} e^{\tau_0 \hat{D}} \right] (x(t_{s+1}), x(t_s))
$$

(35)

We will need to compute the derivatives of this probability with respect to reaction rates:

$$\rho_r \frac{\partial}{\partial \rho_r} \left[ e^{(t_{s+1} - t_s) W} \right] (x(t_{s+1}), x(t_s)) = \sum_{k=0}^{\infty} \int_{t_s}^{t_{s+1}} \cdots \int_{t_s}^{t_{s+1}} d[\tau]_0^n \, \delta\left( (t_{s+1} - t_s) - \sum_{p=0}^{n} \tau_p \right)$$

$$\times \sum_{p=0}^{n} \left[ e^{-\tau_n D} \, \hat{W} \dots e^{-\tau_{p+1} D} \left( \rho_r \, \hat{W}_r \right) e^{-\tau_p D} \dots e^{-\tau_1 D} \, \hat{W} e^{-\tau_0 D} \right] (x(t_s), x(t_s))$$

$$- \sum_{k=0}^{\infty} \int_{t_s}^{t_{s+1}} \cdots \int_{t_s}^{t_{s+1}} d[\tau]_0^n \, \delta\left( (t_{s+1} - t_s) - \sum_{p=0}^{n} \tau_p \right)$$

$$\times \sum_{p=0}^{n} \left[ e^{-\tau_n D} \, \hat{W} \dots \hat{W} (\rho_r \, \tau_p \, D_r) \, e^{-\tau_p D} \, \hat{W} \dots e^{-\tau_1 D} \, \hat{W} e^{-\tau_0 D} \right] (x(t_{s+1}), x(t_s))$$

$$\rho_r [\hat{W}_r]_{IJ} = \left( \frac{\rho_r [\hat{W}_r]_{IJ}}{\sum_r \rho_r [\hat{W}_r]_{IJ}} \right) \left( [\hat{W}]_{IJ} \right) = b_{rIJ} [\hat{W}]_{IJ}$$

where we defined the "branching ratio"

$$b_{rIJ} \equiv \left( \frac{\rho_r [\hat{W}_r]_{IJ}}{\sum_r \rho_r [\hat{W}_r]_{IJ}} \right) = \left\langle \delta_{r, R(I, J)} \right\rangle_{p(I|J)}$$

for reaction $r$ in state $J$, *assuming* each reaction $r$ results in just one output state $I$ per input state $J$. Here $R(I, J)$ is the random variable denoting the actual reaction chosen in transitioning from state $J$ to state $I$. Then

$$\rho_r \frac{\partial}{\partial \rho_r} \left[ e^{(t_{s+1} - t_s) \hat{W}} \right] (x(t_{s+1}), x(t_s)) = \sum_{k=0}^{\infty} \int_{t_s}^{t_{s+1}} \cdots \int_{t_s}^{t_{s+1}} d[\tau]_0^n \, \delta\left( (t_{s+1} - t_s) - \sum_{p=0}^{n} \tau_p \right)$$

$$\times \sum_{p=0}^{n} \left[ e^{\tau_n D} \, \hat{W} \dots e^{\tau_{p+1} D} \left( b_r \, \hat{W} - \hat{W} \rho_r \, \tau_p \, D_r \right) e^{\tau_p D} \dots e^{\tau_1 D} \, \hat{W} e^{\tau_0 D} \right] (x(t_{s+1}), x(t_s))$$

or

$$\rho_r \frac{\partial}{\partial \rho_r} \left[ e^{(t_{s+1} - t_s) \hat{W}} \right] (x(t_{s+1}), x(t_s)) =$$

$$\sum_{k=0}^{\infty} \sum_{p=0}^{n} \left\langle b_r (\text{reaction event } p \text{ out of } n) \right\rangle_{\hat{W}, x(t_{s+1}), x(t_s)} - \rho_r \sum_{k=0}^{\infty} \sum_{p=0}^{n} \left\langle \tau_p \, D_r \right\rangle_{\hat{W}, x(t_{s+1}), x(t_s)}$$

(36)

This finally is a quantity that is easy to compute as a running average during a simulation of the network with incorrect values of the parameters, thereby contributing to the calculation of an improved set of parameter values in a stochastic gradient descent algorithm. This is the key update equation in a learning algorithm for reaction rates in stochastic biochemical networks (extensible to other process networks). Algorithmic details can be found in [18], noting particularly Equation 2.4 therein. A related stochastic learning algorithm is proposed in [8].